\documentstyle[11pt,epsfig]{article}
\newcommand \be{\begin{equation}}
\newcommand \bea{\begin{eqnarray} \nonumber }
\newcommand \ee{\end{equation}}
\newcommand \eea{\end{eqnarray}}
\newcommand{\lp}{\left(}
\newcommand{\rp}{\right)}

\begin{document}

\title{New Evidence of Earthquake Precursory Phenomena\\ in the
17 Jan. 1995 Kobe Earthquake, Japan}

\author{A. Johansen$^1$, H. Saleur$^2$ and D.
Sornette$^{1,3,4}$\\
$^1$ Institute of Geophysics and Planetary Physics,
UCLA, Los Angeles, CA 90095-1567\\
$^2$ Department of Physics, University of
Southern California, Los-Angeles, CA 90089-0484\\
$^3$ Department
of Earth and Space Sciences, UCLA, Los Angeles, CA 90095-1567\\
$^4$ LPMC, CNRS UMR6622 and
Universit\'e de Nice-Sophia Antipolis, 06108 NICE Cedex 2, France}

\maketitle

\begin{abstract}

Significant advances, both
in the theoretical understanding of rupture processes in heterogeneous
media and in the methodology for characterizing critical behavior,
allows us to reanalyze the evidence for criticality and especially
log-periodicity
in the previously reported chemical anomalies that preceded the Kobe
earthquake.
The ion ($Cl^-$, $K^+$, $Mg^{++}$, $NO_3^{-}$ and $SO_4^{--}$)
concentrations
of ground-water issued from deep wells located near the
epicenter of the 1995 Kobe earthquake are taken as proxies for the
cumulative damage
preceding the earthquake. Using both a parametric and non-parametric
analysis,
 the five data sets are compared extensively to synthetic time series.
The null-hypothesis that the patterns documented on these times series
result from noise decorating a simple power law is rejected with a very
high
confidence level.

\end{abstract}

\pagebreak

Since the inception of seismology, the search for reliable precursory
phenomena
of earthquakes has shown that the seismic process is preceded by a
complex
set of physical precursors. In addition to
the direct seismic foreshocks \cite{Abercrombie,Maeda} and seismic
precursors
\cite{Knopoff,Bowmanetal}, many other
anomalous variations of various geophysical variables such as electric
and
magnetic fields and conductivity \cite{electric,debatevan},
as well as chemical concentration \cite{Tsunogai,Igarashi} have been
documented.
However, there is no consensus on the
statistical significance of these precursors and their reliability
\cite{Wyss,Geller},
due to a lack of reproducibility and of understanding of the underlying
physical
mechanisms.

One of the most striking reported seismic precursory phenomena are the
time dependence of ion concentrations of ground-water issuing from
deep wells located near the epicenter\cite{Tsunogai} and the
ground water radon anomaly \cite{Igarashi}
preceding the earthquake of magnitude $6.9$ near Kobe, Japan, on January
17, 1995.
Since the first quantitative analysis of this data, which suggested
a discrete scale-invariant time-to-failure behavior \cite{Johansenetal},
significant advances, both
in the theoretical understanding of rupture processes in heterogeneous
media and in the methodology needed to characterize critical behavior,
permits a
reassessment of the data.

Within the critical earthquake model \cite{SorSam,Bowmanetal}, a large
earthquake is viewed as the culmination of a cooperative organization of
the stress and of the damage of the crust in a large area extending over
distances
several times the size of the seismic rupture \cite{Knopoff}. Most of
the
recently
developed mechanical models
\cite{thermalfuse,Newmandyn2,Sahimi,Zapperi,sdZhang}
and experiments \cite{Anifrani,Ciliberto}
on rupture in strongly heterogeneous media
(which is the relevant regime for the application to the earth)
view rupture as a singular `critical' point \cite{tricrit}:
the cumulative damage $D$, which can be measured by acoustic emissions,
by the total number of broken bonds or by the total surface of new
rupture
cracks,
exhibits a diverging rate as the critical stress
$\sigma_c$ is approached, such that $D$ can be written as an
`integrated susceptibility'
\be
D \approx A+ B(\sigma_c-\sigma)^{z} ~,   \label{eaka}
\ee
 The critical exponent $0 < z < 1$ is equal to $1/2$ in mean field
theory
\cite{Sor2,Zapperi}
and can vary depending on, {\it e.g.}, the coupling corrosion and healing
processes.
In addition, it has been shown \cite{Anifrani,DLA,canonical}
that log-periodic corrections decorate the leading power law behavior
(\ref{eaka}),
as a result of intermittent amplification processes during the rupture.
They have also been suggested for seismic precursors \cite{SorSam}.
This log-periodicity introduces a hierarchy of characteristic
time
and length scales with a prefered scaling ratio $\lambda$
\cite{DSI}. As a result, expression (\ref{eaka}) is modified into
\be
D \approx A + B \left(\sigma_c - \sigma \right)^z +
  C \left(\sigma_c -\sigma \right)^z \cos\left( 2\pi f \ln
\left(\sigma_c
-\sigma\right) +
    \phi \right)~ ,  \label{pwlg}
\ee
where $f = 1/\ln(\lambda)$. Empirical \cite{Anifrani}, numerical
\cite{DLA,canonical} as well as theoretical analyses \cite{DSI}
point to a prefered value $\lambda \approx 2.4 \pm 0.4$, corresponding
to
a frequency $f \approx 1.2 \pm 0.25$ or radial frequency $\omega = 2 \pi
f
\approx 7.5 \pm 1.5$. A value for $\lambda$ close to $2$ is suggested on
general grounds
from a mean field calculation for an Ising and, more generally, a Potts model
on a hierarchical lattice in the limit of an infinite number of
neighbors
\cite{DSI}.  It also
derives from the mechanisms of a cascade of instabilities in competing
sub-critical
crack growth \cite{DLA}. Empirically, we see that there is not rarely a bias
towards a
value twice this, corresponding to a better of signal-to-noise ratio for
rescalings of $\lambda^2$ in the underlying renormalization group equation

The physical model underlying our analysis posits that the measured ion
($Cl^-$, $K^+$, $Mg^{++}$, $NO_3^{-}$ and $SO_4^{--}$) concentrations
of ground-water issued from deep wells located near the
epicenter of the 1995 Kobe earthquake are proxies for the cumulative
damage
preceding the earthquake.
In this reaction-limited model, any fresh rock-water interface created
by the
increasing damage leads to the dissolution of ions in the carrying fluid
that can be detected in the wells. We thus expect that the time
evolution
of measured ion concentration should follow closely the equation
(\ref{pwlg}).
Due to the large heterogeneity of rocks, this `deterministic' signal
should be decorated by noise with different realizations for each
ion originating from different rock types. To test our
hypothesis, we analyze the five ion data sets, thus increasing
the statistical significance over our previous report
\cite{Johansenetal}.
However, it is possible that their noise realizations are not completely
independent as some of the anions and cations are
necessarily coupled pairwise. As we shall see, the poor result obtained
from
the analysis of $K^+$ might be due to the fact that it is generally
coupled to both $Cl^-$ and $NO_3^-$.

\begin{figure}
\parbox[l]{0.45\textwidth}{
\epsfig{file=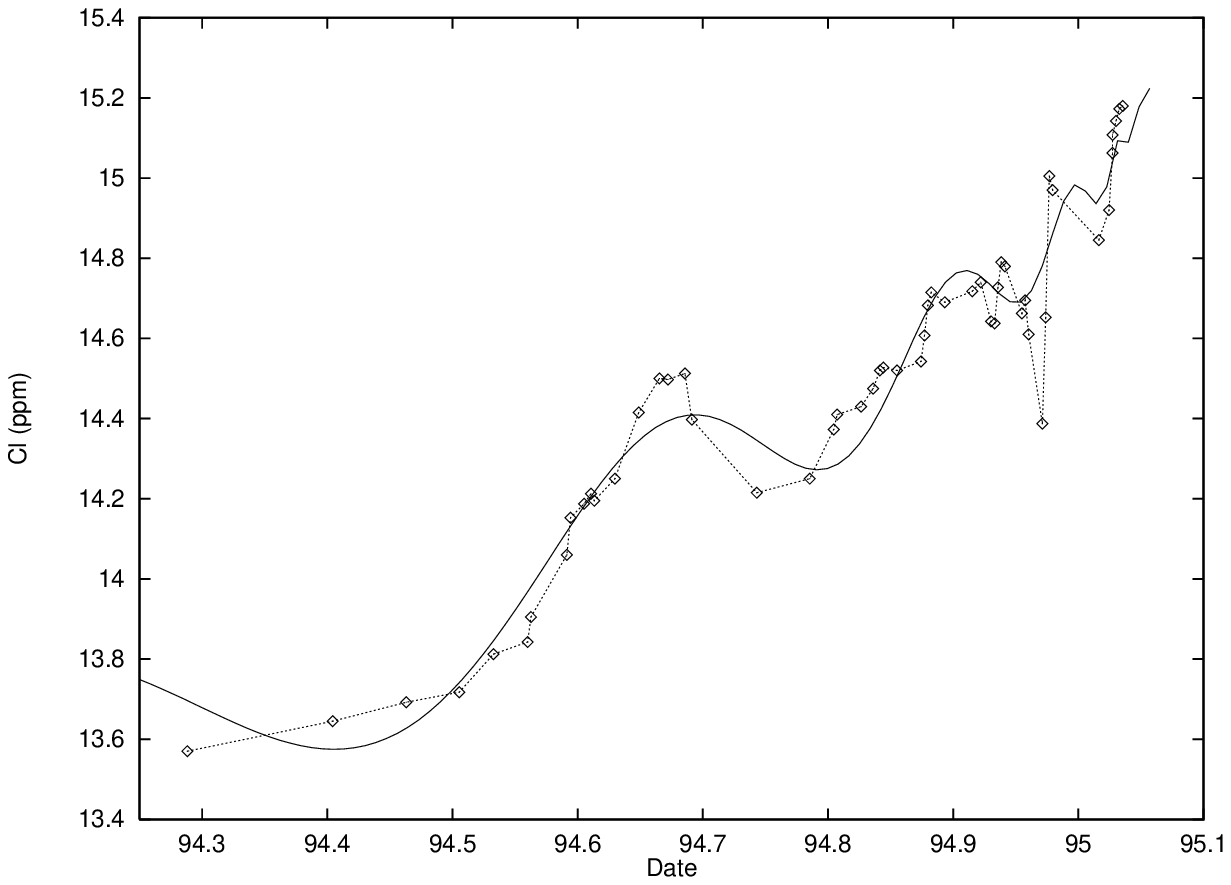,width=0.45\textwidth}}
\parbox[r]{0.45\textwidth}{
\epsfig{file=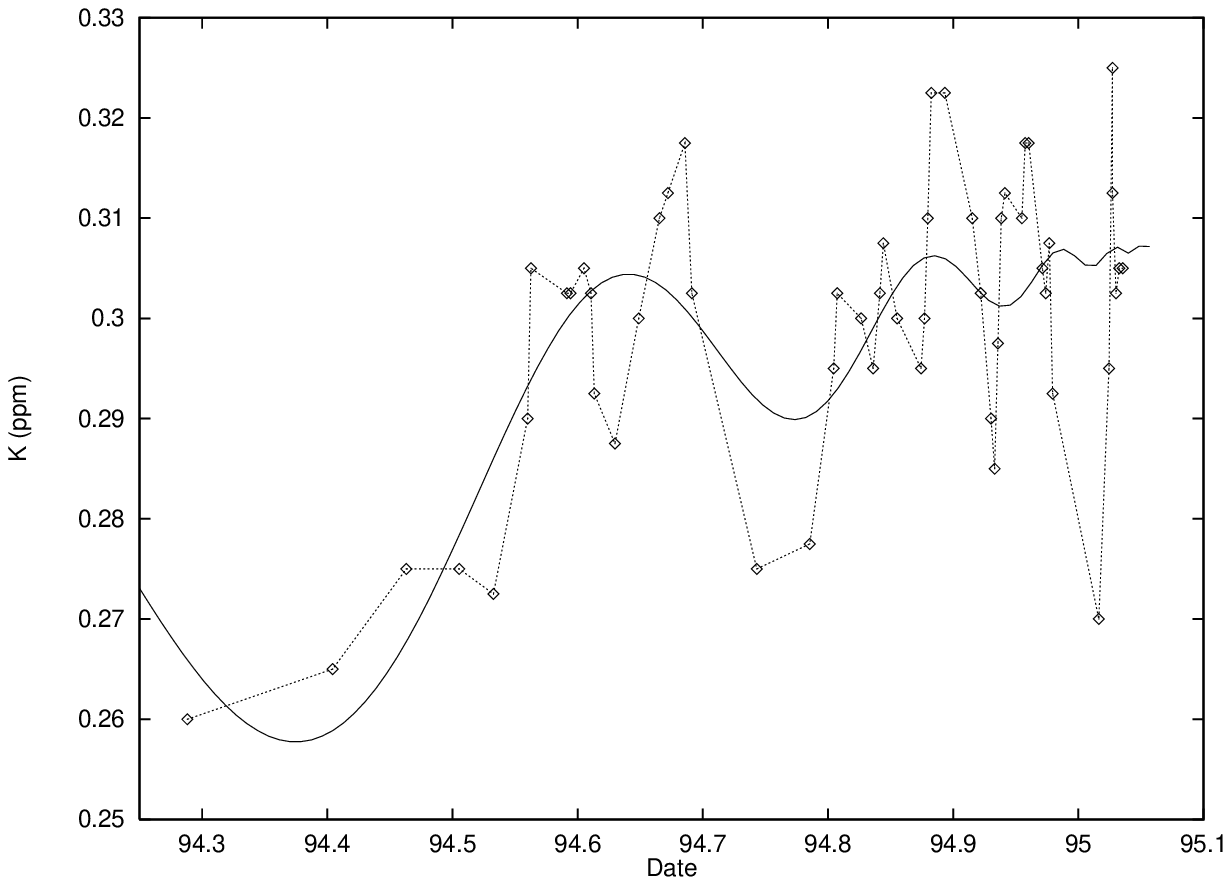,width=0.45\textwidth}}
\parbox[l]{0.45\textwidth}{
\epsfig{file=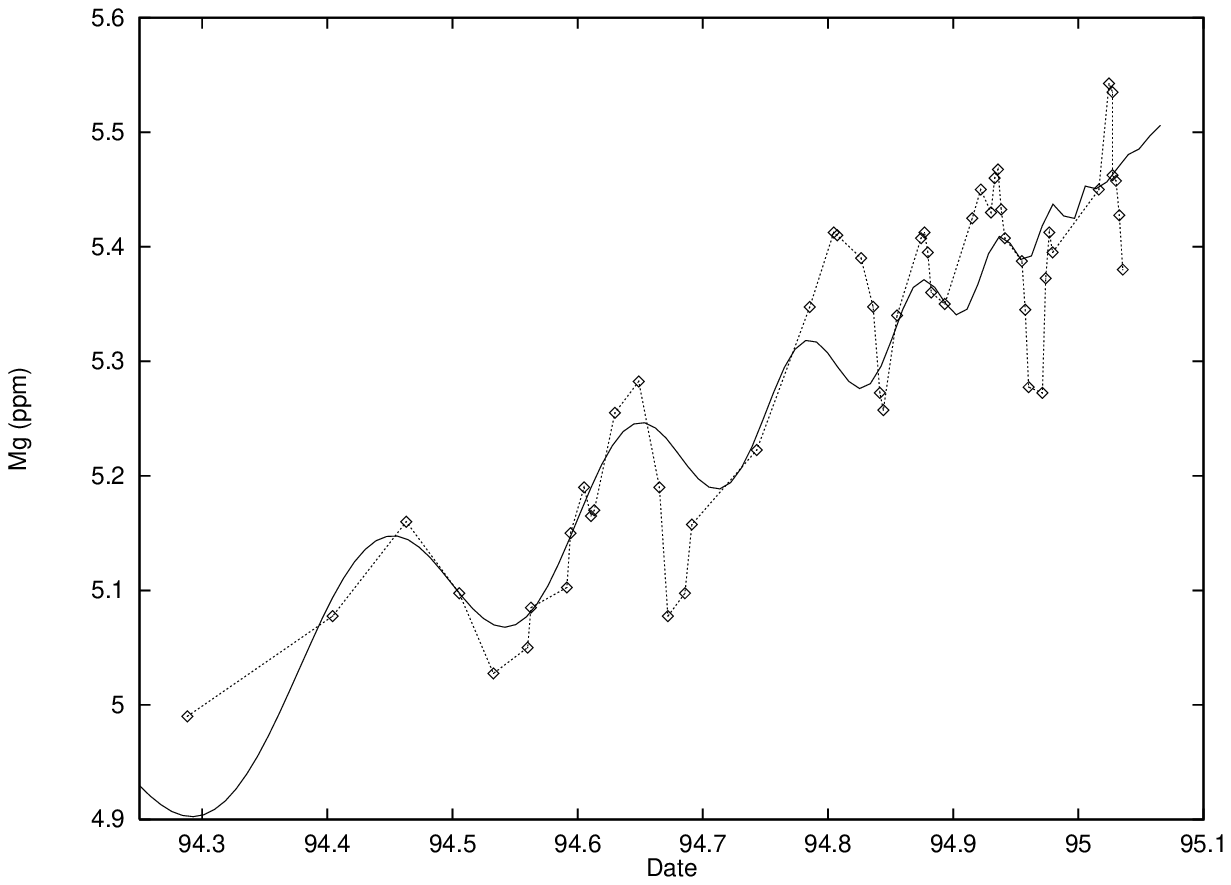,width=0.45\textwidth}}
\parbox[r]{0.45\textwidth}{
\epsfig{file=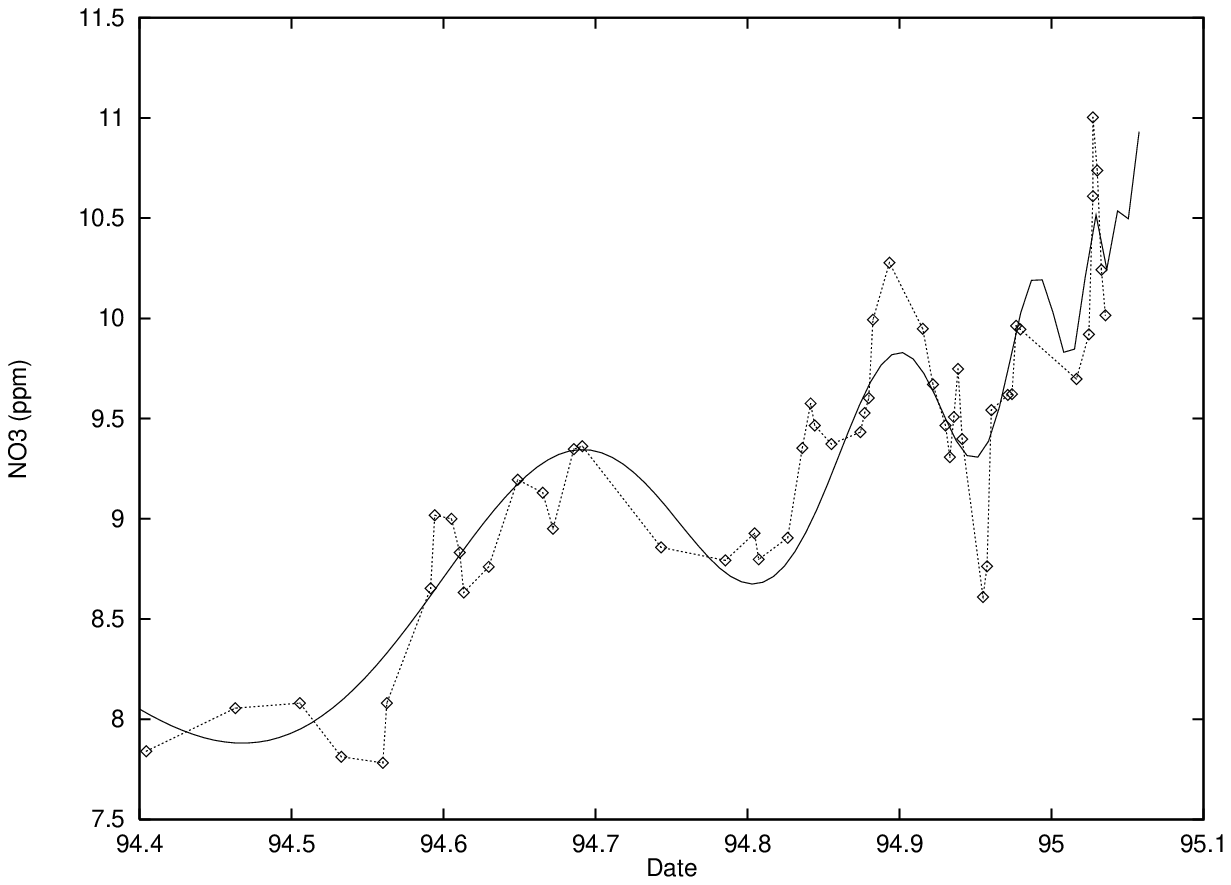,width=0.45\textwidth}}
\parbox[l]{0.45\textwidth}{
\epsfig{file=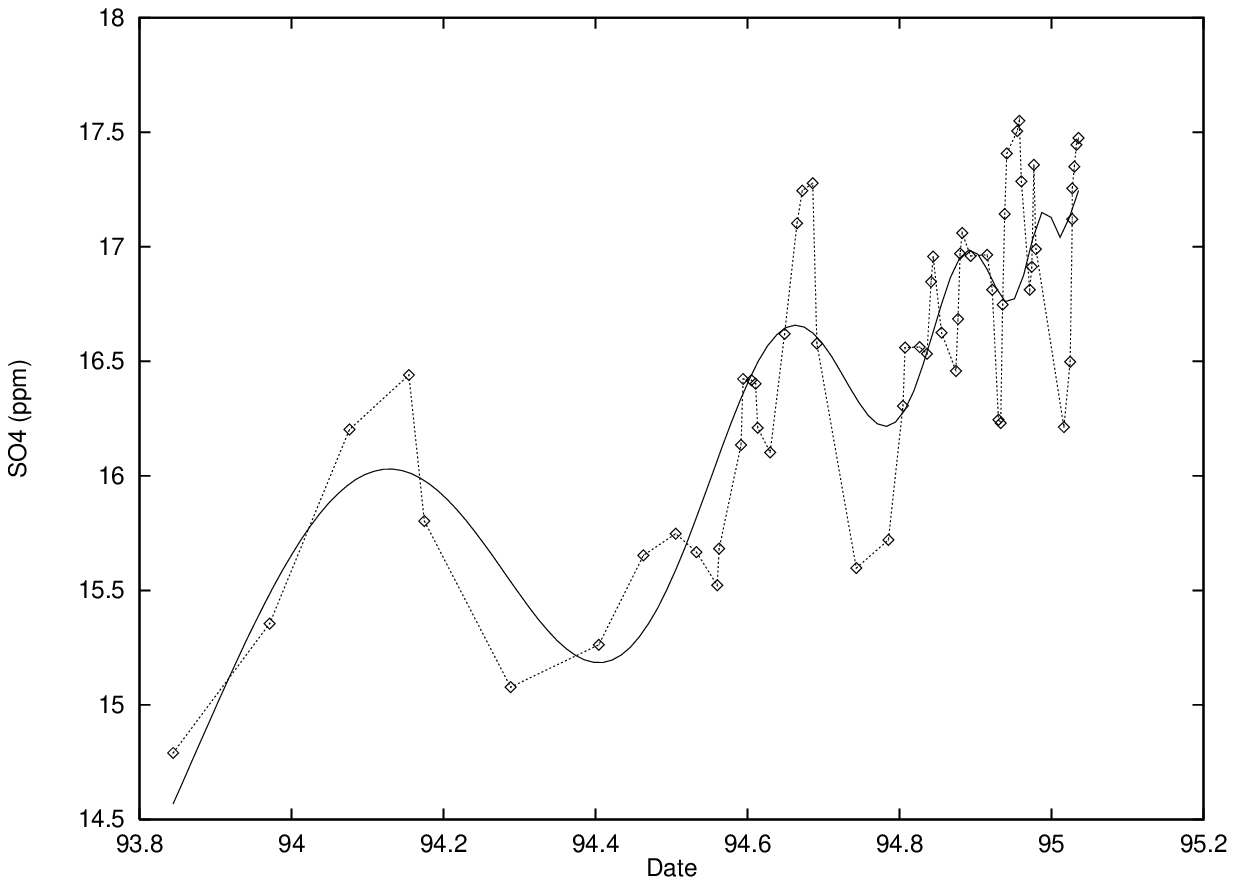,width=0.45\textwidth}}
\hspace{15mm}
\parbox[r]{0.45\textwidth}{
\epsfig{file=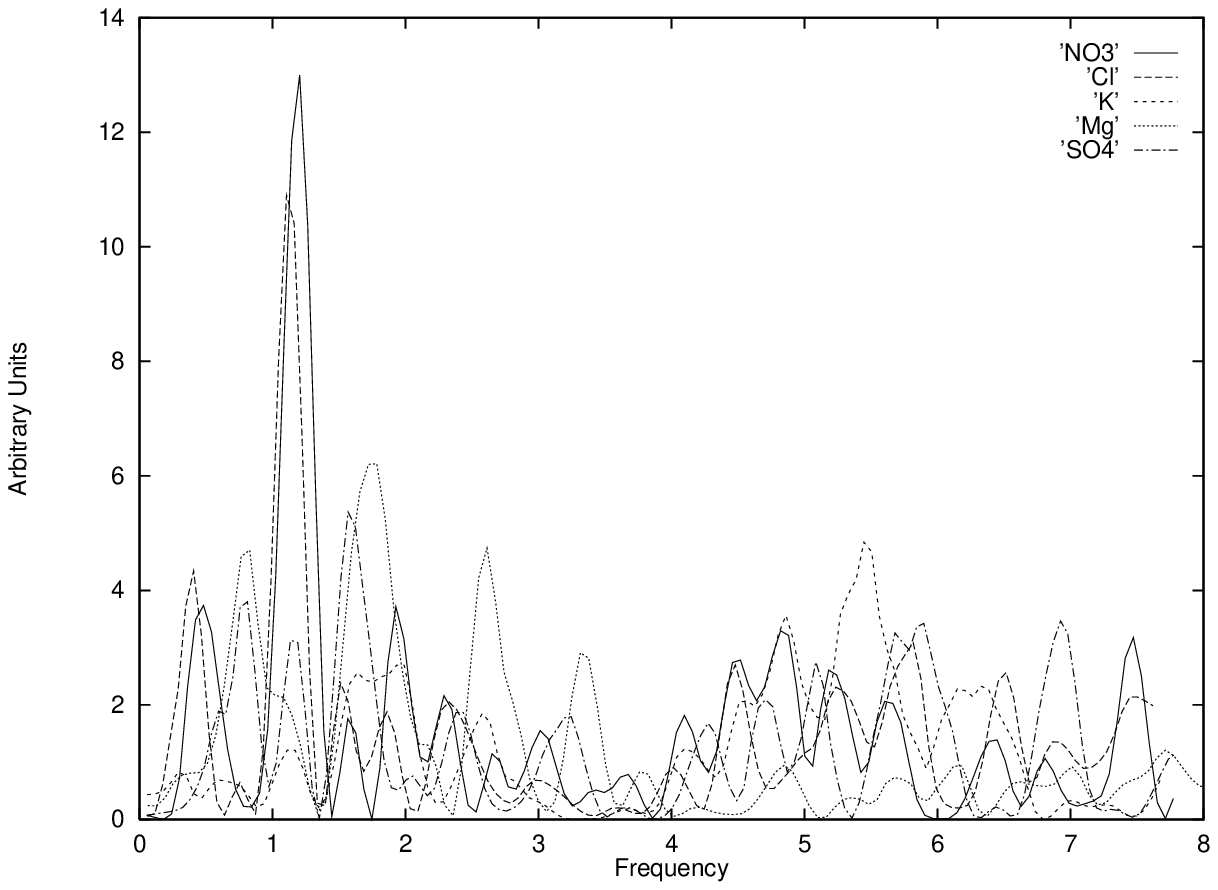,width=0.45\textwidth}}

\caption{\protect\label{kobefits}
The concentrations of
$Cl^-$ (top left), $K^+$ (top right), $Mg^{++}$ (middle left),
$NO_3^{-}$
(middle right) and $SO_4^{--}$ (lower left) as a function of date.
The fit is equation (\protect\ref{pwlg}). The values of the fits are
$Cl^-$ : $A\approx 15.3$, $B\approx -1.9$, $C\approx 0.34$, $z\approx
0.58$,
$t_c\approx 95.060$, $\phi\approx -0.40$ and $\omega \approx  7.0$.
$K^+$: $A\approx 0.31$, $B\approx -.044$, $C\approx   0.036$, $z\approx
1.2$,
$t_c\approx  95.061$, $\phi\approx 0.05$ and $\omega \approx  7.2$.
$Mg^{++}$: $A\approx 5.5$, $B\approx -0.65$, $C\approx 0.11$, $z\approx
0.82$,
$t_c\approx 95.069$, $\phi\approx -1.3$ and $\omega \approx 16$.
$NO_3^{-}$: $A\approx 11.6$, $B\approx -3.7$, $C\approx  -0.67$,
$z\approx
0.28$,
$t_c\approx 95.059$, $\phi\approx 2.1$ and $\omega\approx 7.5$.
$SO_4^{--}$: $A\approx 17.3$, $B\approx  -2.2$, $C\approx  0.83$,
$z\approx
0.77$,
$t_c\approx  95.066$, $\phi\approx  -.31$ and $\omega \approx  7.4$.}
Lower right figure is the Lomb periodogram of the five time series
as a function of log-frequency defined as the conjugate to
$\log\left( \left( t_c - t\right) /t_c\right)$. Prior
to the spectral analysis, the leading power law has been removed using
equation (\ref{residue}). The values of the Lomb peaks are ~11, ~3,
~6, ~13 and ~5 in the same order as above.

\end{figure}

Figure \ref{kobefits} shows the five data sets on which a moving average
using three
points has been applied. In this moving average, the middle point
as usual carries double weight except for the two endpoints. Each of the
five data
sets is fitted with equation (\ref{pwlg}).
The time intervals used in the fit were determined consistently for
all five data sets by identifying the date of the lowest value of the
concentration and using that date as the first data point. The last
data point was taken as the last measurement before the earthquake.
This gave $53$ points for the first three data sets, 52 for the fourth
and
58 points for the fifth data set. For three of the data sets
($Cl^-$, $K^+$, and $NO_3^{-}$), the fit shown is also the best fit. In
the case of $SO_4^{--}$, the best fit had an
angular log-frequency $\omega \approx 19$ well above the expected range
$6 \leq \omega \leq 9$, while the second best fit has $\omega \approx 7$
within the expected range and is shown instead.
This is also the case for $Mg^{++}$, where the
best fit has a very low angular log-frequency $\omega \approx 2$
which capture nothing but a slowly varying trend. Here, the angular
log-frequency of the second best fit $\omega \approx 16$ is
approximately double of what is found for the other four data sets
corresponding to a squaring of $\lambda$ as previously discussed. We
note that the ranking of the fits is done using the variance between the
data and the fit with equation (\ref{pwlg}), where the stress $\sigma$
has
been replaced with time $t$, $t_c$ being the critical date of the
earthquake.
This assumes a Gaussian white noise-distribution, which constitutes a
reasonable null-hypothesis for the noise but presumably does not fully
reflect the reality.

We observe that the values of the critical time $t_c$ predicted
for the earthquake from the
various fits are not only very stable but also remarkably consistent
with the true date $95.053$ of the earthquake. And except for the case
of
$Mg^{++}$, where it has approximately doubled,
the same is true for the angular log-frequency $\omega$ (although in
2 cases, we took the second best fit as explained above).  This
robustness with respect to the date of the earthquake and the preferred
scale ratio $\lambda = e^{2\pi / \omega} \approx 2.5$ is quite
remarkable,
considering that the value obtained for the exponent $z$ varies by
approximately a factor of $4$ from the smallest to the largest value.

We complement these fits by a direct `non-parametric' analysis
of the log-periodic component, taken as
a crucial test of the critical earthquake model captured
by (\ref{eaka}), (\ref{pwlg}).
After de-trending each of the five chemical time series $c(t)$ using
\be \label{residue}
c\lp t\rp \rightarrow \frac{c\lp t\rp -
\left[ A + B (t_c-t)^{z}\right]}{C(t_c-t)^{z}}~,
\ee
which should leave a pure log-periodic cosine if no other effects were
present,
we apply a Lomb
periodogram \cite{Lomb} to the de-trended data as a function of
$\log\lp \lp t_c - t\rp /t_c\rp$.
Lower fight figure of \ref{kobefits} shows the five spectra
superimposed.
In all case except for $K^+$, we observe significant peaks with a
log-frequency
$f\equiv {\omega \over 2\pi}$ between $1$ and $2$, i.e.
$\omega$ is approximately between $6$ and $12.5$, and in two cases
($Cl^-$ and $NO_3^-$), we
have two very clear peaks at $f \approx 1.2$ ($\omega \approx 7.5$)
standing out
with a Lomb weight of $\approx 11$ and $\approx 13$, respectively, in
agreement
with our prediction.

In order to assess the significance of these results, we present
rather exhaustive statistical tests performed
by constructing synthetic time series that differ from the real data
only by the log-periodic pattern and we of course follow the same
testing
procedure.

There are several reasons why the results from the analysis above
should be compared with those of synthetic tests. In particular,
the real time series have been sampled non-uniformly in time because the
water from
the wells was collected from commercial bottles of which the production
dates could be identified. We observe that the depletion process of bottles
in stores implies that the number of bottles with a production date $t$
prior to the date of the earthquake $t_c$ is proportional to
$\log \lp t_c - t\rp$, leading to a uniform sampling in
$\log \lp t_c - t\rp$ instead of time $t_c - t$ to the earthquake.
In order to investigate the effect of this sampling of the Kobe data
with respect to log-periodic signatures, we have as a first step
generated
twenty
synthetic data sets each with $56$ points as in the original data
analyzed in
ref.\cite{Johansenetal}. The synthetic data was generated using a noisy
power law with the
parameter values of the leading power law of the fit presented in
ref.\cite{Johansenetal}.
Specifically, the equation
$c\lp t_i\rp = 15.45 + \left( -1.877 + k\cdot \lp 0.5-\mbox{ran}\rp
\right)
\cdot
\lp 95.053 - t_i \rp^{0.44}$
was used with different noise-levels $k$. Here, ran is a uniform
random number generator with values in the interval $\left[ 0:1 \right[$
from
\cite{Press}.
In order to obtain a sampling similar to the one found in the
original data, the sampling times $t_i$ of the synthetic data was chosen
as
$t_i= 95.053 - (95.053 - 93.847)e^{-5 \mbox{ran}}$,
where $93.847$ was the date of the first point in the original analysis.
This
then gives us a distribution of sampling times, which are uniform in
log-time to $t_c$, where $t_c = 95.053$ is the time at which the Kobe
earthquake occurred. As noise-level, we used $k=0$, $k=0.37$ and
$k=0.73$, corresponding to no-noise, a noise level of the same amplitude
as the estimated log-periodic oscillations and a noise level twice that.

In the case of no noise $k=0$, the Lomb Periodograms exhibit in general
a forest of peaks over a large log-frequency interval which
average out to give a flat spectrum.
Adding noise of amplitude $k=0.37$ and $k=0.73$ enhances the peaks
but the averaging over the
periodograms of the different synthetic data sets again produces an
essentially flat spectrum. By ``flat'', we mean that, in each individual
synthetic time series, about $7$ ``peaks'' are found
above $70$\% of the largest one over the entire log-frequency interval
$0 <
f < 6$.
This suggests a qualifier for the significance of a Lomb peak,
corresponding to
counting the number of crossings between the periodogram and a
$70$\%-level or a $50$\%-level, taking the highest peak as the $100$\%
reference (a single peak corresponds to two crossings). For all cases,
we
obtain at least $10$ crossings depending on how
one defines a peak. We also see that the noise has the effect that the
largest peak is no longer to be found for the lowest frequencies.

We now prodeed with synthetic tests.
As a first null-hypothesis, we take white noisy power law.
An indication of the statistical significance of the results obtained
from the chemical data can be estimated by comparing the periodogram
of the de-trended data (\ref{residue}) with the periodograms of a
sequence of random numbers uniformly distributed between $0$ and $1$
after a moving average over three points
has been applied. Hence, we have generated $1000$ sequences with
approximately the same number of points ($53$) and the time interval of
the true data\,: none had a Lomb peak above 10.  Out of $10.000$
synthetic data sets, 9 had Lomb peaks above 10, but none above 12. An
additional $100.000$ data sets were generated providing three Lomb
peaks above 12, but none of them had a log-frequency between $1$ and
$2$. For peak values above 10, the $100.000$ data set had six such peaks
in that log-frequency range.
If we use a Gaussian white noise distribution instead of the uniform
one,
out of $1000$ synthetic
data sets, we get $20$ peaks with a log-frequency between $1$ and $2$
and a Lomb weight above $6$, but none above $10$ with a log-frequency
between $1$ and $2$. Out of $10.000$ data sets, only one such peak is
observed.

Many other systematic and more elaborate tests have been performed with and
without smoothing, as well
as by directly
generating noisy power laws and then performing the fit with
(\ref{pwlg})
and the de-trending with (\ref{residue}). We find that smoothing after
de-trending is not significantly different from smoothing before
de-trending. Similarly, a generation of noisy power laws gives results
similar to the foregoing white noise hypothesis.

\begin{figure}
\begin{center}
\epsfig{file=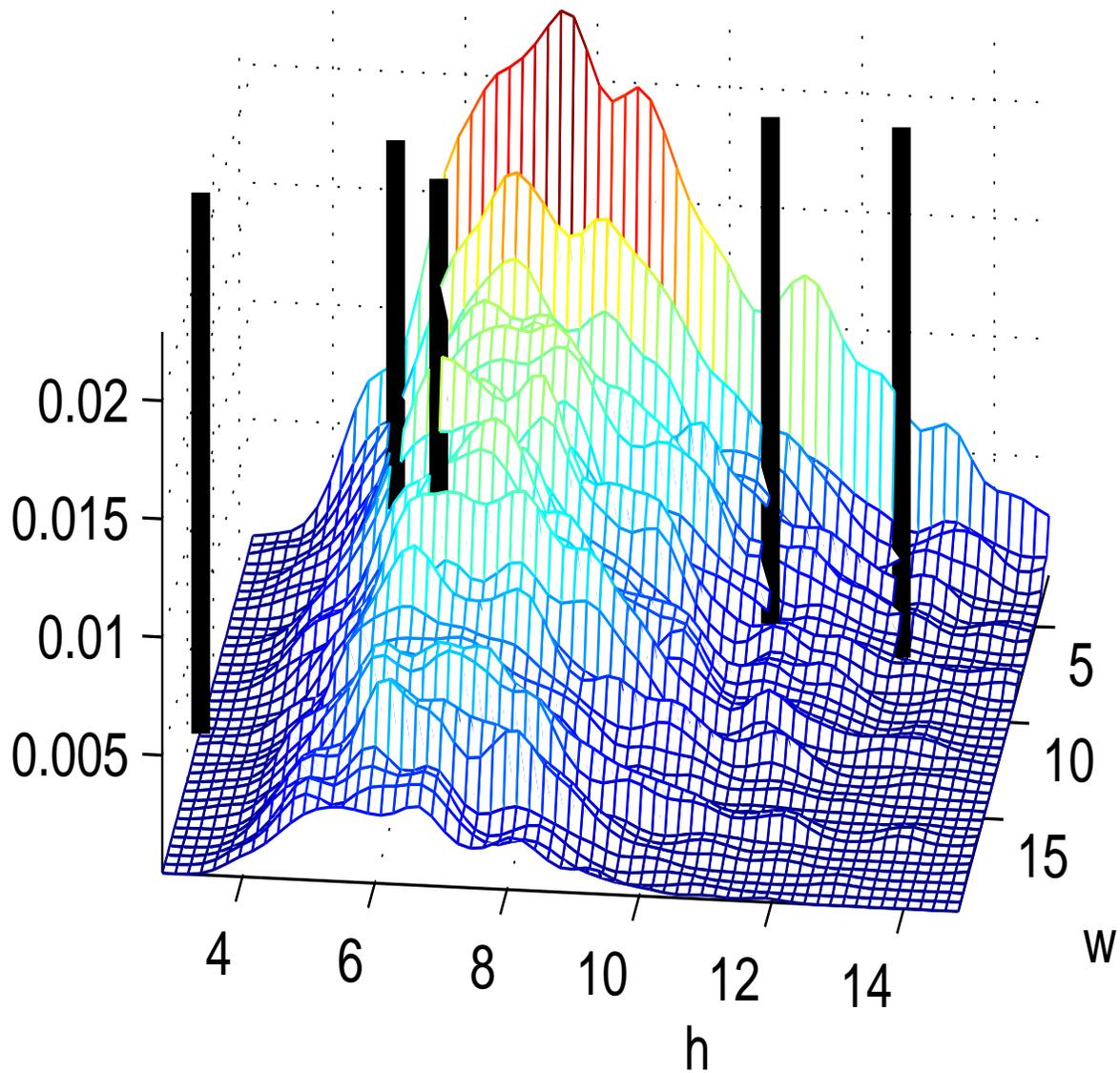,height=16cm,width=16cm}
\caption{\protect\label{fig2} Bivariate distribution for the null hypothesis
of angular log-frequencies $\omega$ and Lomb peak
$h$ obtained from a series of 100,000 synthetic time series.
The five vertical lines correspond to the five real times series. They
are obtained from the lower right panel in figure 1 by locating the
log-frequency and Lomb peak height of the highest peak for each data set.
}
\end{center}
\end{figure}

Figure \ref{fig2} presents a global summary of our tests by showing the
bivariate distribution of angular log-frequencies $\omega$ and Lomb peak
$h$ obtained
from a series of 100,000 synthetic time series.
The five vertical lines correspond to the five real times series and can
be
obtained from the lower right panel in figure 1 by locating the
log-frequency
and Lomb peak height of the highest peak for each data set. One can
visualize
that not only the fundamental frequency but its harmonics are sometimes
visible
($Cl^-$ and $NO_3^{-}$) and, as we said above, may be dominating
($Mg^{++}$
and $SO_4^{--}$).

The interpretation of these results is that there is  a confidence
level of $99.99$ \% for a single peak with a log-frequency between
$1$ and $2$ and a Lomb weight above $10$. This confidence
 interval is evaluated with respect to our initial null hypothesis of
of uncorrelated (white) noise and would of course
change with a different null hypothesis. Nevertheless, the confidence
level achieved with the present null-hypothesis is above 99.9\% and we hence
expect it to remain would remain very high over a range of null-hypothesis.
Also, the confidence level will also change depending on whether one
assumes that the various ion concentrations are independent or not.

In conclusion, we wish to stress that the presented analysis constitute
but a single case-study.  Hence, it does not propose a
recipe for earthquake prediction. However, we feel that the statistical
evidence for this particular analysis is significant enough to encourage
further studies along similar lines.

\smallskip
\noindent {\bf Acknowledgments}: We are grateful to H. Wakita for kindly
providing the data and for useful correspondence.
H.S. thanks Y. Huang for collaboration
at an early stage of this work and for many useful discussions. For
further discussion of statistical tests and interpretations from a
different point of view, see \cite{yueqiang}.

\end{document}